\documentclass[%
  twocolumn,%
  aps,%
  superscriptaddress,
  showpacs,%
  letterpaper,%
  amsfonts,%
  footinbib,%
  floatfix%
]{revtex4}

\usepackage[pdftex]{graphicx}
\usepackage{amsmath,amssymb,bm}
\usepackage{hyperref}
\hypersetup{%
  breaklinks = {true},
  citecolor = {blue},
  colorlinks = {true},
  linkcolor = {red},
  pdfauthor = {\textcopyright\ Vitor M. Pereira },
  pdfcreator = {\LaTeX\ and \flqq hyperref\frqq},
  pdffitwindow = {true},
  pdfmenubar = {true},
  pdfpagelayout = {SinglePage},
  pdfstartview = {Fit},
  pdftoolbar = {true},
  plainpages = {false},
}

\newcommand{\Ocal}{\ensuremath{\mathcal{O}}}

\newcommand{\Tcal}{\ensuremath{\mathcal{T}}}
\newcommand{\bd}{\ensuremath{\bm{\delta}}}

\newcommand{\bn}{\ensuremath{\bm{n}}}

\newcommand{\bk}{{\ensuremath{\bm{k}}}}

\newcommand{\bp}{{\ensuremath{\bm{p}}}}
\newcommand{\bq}{{\ensuremath{\bm{q}}}}

\newcommand{\bR}{{\ensuremath{\bm{R}}}}

\newcommand{\ce}{{\ensuremath{\varepsilon}}}
\newcommand{\e}{\epsilon}

\newcommand{\Fref}[1]{fig.~\ref{#1}}
\newcommand{\Eqref}[1]{eq.~\eqref{#1}}
\newcommand{\Ref}{ref.~}
\newcommand{\zdir}{\ensuremath{\mathcal{ZZ}}}
\newcommand{\adir}{\ensuremath{\mathcal{AA}}}

\begin{document}


\title{Optical Properties of Strained Graphene}

\author{Vitor M. Pereira}
\affiliation{Graphene Centre and Department of Physics, 
National University of Singapore, 2 Science Drive 3, Singapore 117542}

\author{R. M. Ribeiro}
\affiliation{Centro de F\'{\i}sica  e  Departamento de
F\'{\i}sica, Universidade do Minho, P-4710-057, Braga, Portugal}

\author{N. M. R. Peres}
\affiliation{Centro de F\'{\i}sica  e  Departamento de
F\'{\i}sica, Universidade do Minho, P-4710-057, Braga, Portugal}

\author{A.~H. Castro Neto%
  \footnote{On leave from Physics Department, Boston University.}
}
\affiliation{Graphene Centre and Department of Physics, 
National University of Singapore, 2 Science Drive 3, Singapore 117542}

\pacs{78.67.Wj, 81.05.ue, 73.22.Pr}


\date{\today}

\begin{abstract}
The optical conductivity of graphene strained uniaxially is studied
within the Kubo-Greenwood formalism. Focusing on inter-band
absorption, we analyze and quantify the breakdown
of universal transparency in the visible region of the spectrum,
and analytically characterize the transparency as a function of strain
and polarization. 
Measuring transmittance as a function of incident polarization
directly reflects the magnitude and direction of strain. 
Moreover, direction-dependent selection rules permit 
identification of the lattice orientation by monitoring the
van-Hove transitions. These photoelastic effects in graphene
can be explored towards atomically thin, broadband optical elements.
\end{abstract}

\maketitle


%
%
\section{Introduction}
Transparent flexible electronics is currently a much sought
technology,
with applications that can range from foldable displays and electronic
paper, to transparent solar cells. Graphene, on account of
its ultimate thickness, large transparency \cite{Nair:2008}, high
mechanical resilience under strong stress/bending cycling,
and excellent electronic mobility \cite{Bolotin:2008}, has been
promptly ranked among the best placed materials to achieve those
technologies \cite{Ruoff:2009,Bae:2010}. 
Such goals require a
thorough understanding of the interplay between the different aspects
that will unavoidably be present in such devices, namely how sensitive
the dielectric and optical properties of graphene are to gating and
straining. At the same time, broadband optical elements that can be
scaled down to the nanoscale, and easily integrated into
photonic/photoelectronic circuits, are equally appealing scenarios in
current nanotechnology.

The optical absorption response of graphene has been recently given
thorough attention on both the experimental
\cite{Nair:2008,Basov:2008,Heinz:2008,Kuzmenko:2008,Heinz:2010} and
theoretical fronts
\cite{Ando:2002,Peres:2006,Falkovsky:2007,Mischenko:2009,Sheehy:2009,
Gusynin:2009,Peres:2010,RMP:2009,Peres-RMP:2010}. 
One of the distinguishing features of undoped (or lightly doped)
graphene arises from the constancy of its transparency,
$\Tcal(\omega)$, which is controlled by electron-hole excitation
processes, and universal: 
$\Tcal(\omega) \approx 1 - \pi \alpha$ ($\alpha\simeq 1/137$ being the
fine
structure constant) \cite{Nair:2008}.
This universality is a consequence of the particle hole symmetry of
graphene's spectrum, combined with the cancellation of the
frequency dependencies of the matrix element and vanishing density of
states. It holds throughout a broad
spectral range comprising the frequencies between the Fermi energy,
$\mu$, and the vicinity of the van~Hove singularity (VHS) at the $M$
point in the Brillouin zone (BZ). 
In undoped graphene this covers a band spanning the visible region,
down to the far IR.

Here we analyze and quantify how the optical absorption associated
with direct optical transitions is affected by strain. The
strain-induced anisotropy leaves a clear signature in the optical
response of the system, modulating its transmittance, reflectance and
absorption, while simultaneously rotating the polarization of incoming
light. Tailoring of graphene systems on the basis of such properties
extends the concept of strain engineering in graphene from
electronic structure and transport \cite{Pereira:2009a,Guinea:2010}
to the optical domain as well. Such effects are described next.
Related work on effects of strain in graphene's optical response has
been reported in
Refs.~\cite{Pellegrino:2009,Pellegrino:2010,Sinner:2010}.

\begin{figure}[bt]
  \centering
  \includegraphics[width=0.45\textwidth]{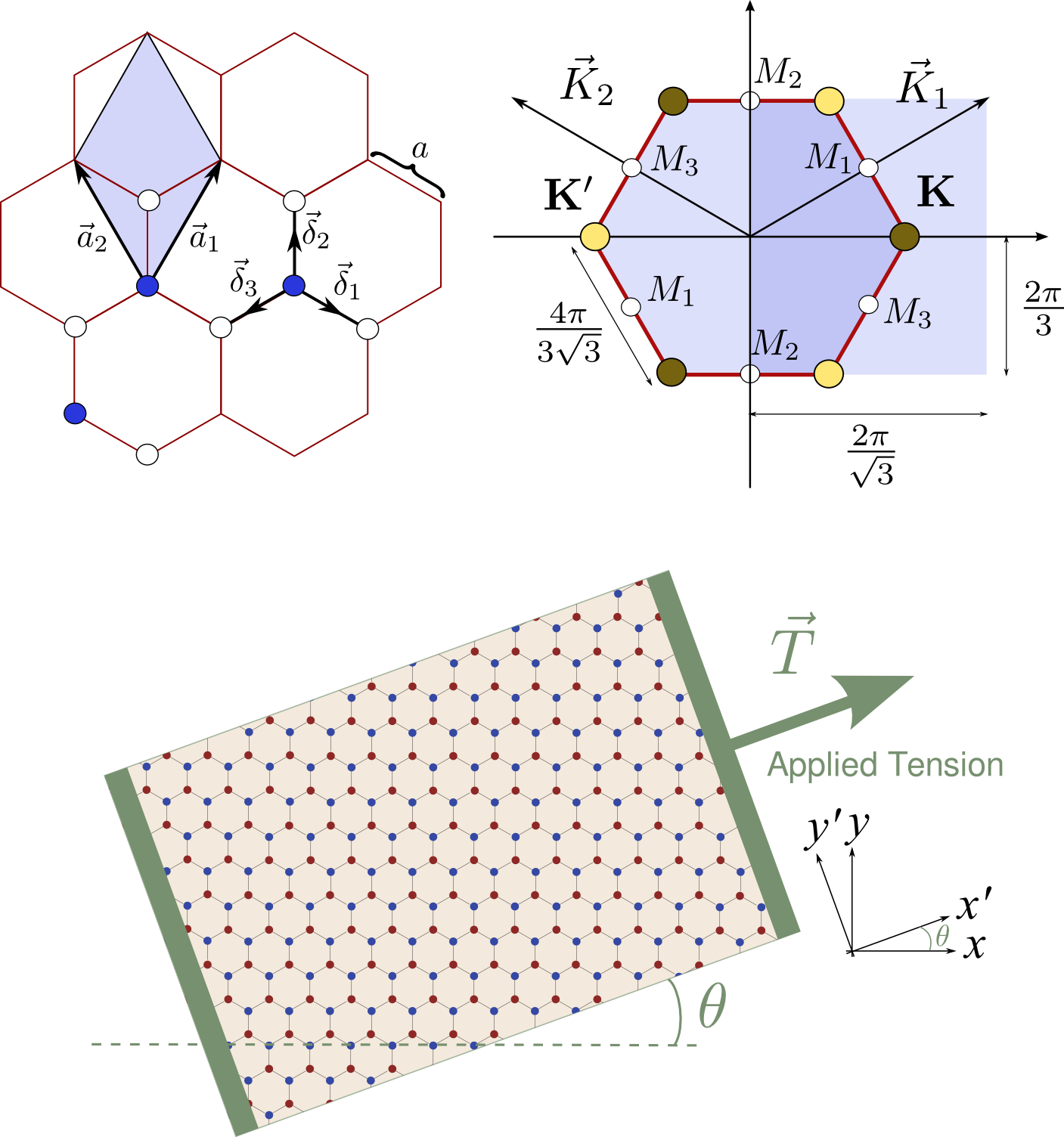}
  \caption{
    Lattice orientation in direct and reciprocal space,
    overlaid with several quantities used in the text.
    The bottom illustration shows the relative orientation of the 
    lattice with respect to a general tension direction $\theta$.
  }
  \label{fig:Fig-1}
\end{figure}

%
%
\section{Strain-Induced Anisotropy}
We concentrate on uniform uniaxial deformations of free
graphene, which provide the highest degree of anisotropy for given
amount of strain, $\ce$. Within a tight-binding approximation for the
$\pi$-band subsystem, strain impacts electronic motion via the
modification of the Slater-Koster parameters $t_i \equiv t(\bd_i)$ in
the Hamiltonian
\begin{equation}                                                      
  H = \sum_{\bR,\bd} t(\bd) \,
    a^\dagger_{\bR}b^{\phantom{\dagger}}_{\bR+\bd} 
    + \text{H. c.}       
  \,.                                                                 
  \label{eq:Hamiltonian}                                              
\end{equation}                                                        
Here $\bR$ denotes a position in the Bravais lattice; $\bd_{1,2,3}$
connects site $\bR$ to its neighbors; $a(\bR)$ and $b(\bR)$ are the
field operators in sublattices A and B. In what follows, we
characterize the interplay between strain and electronic structure in
the same framework described in detail in
\Ref~\onlinecite{Pereira:2009b}. In summary, 
this entails the assumption that hopping amplitudes vary with
distance as 
$t(r) \simeq t e^{-\lambda(r-a)}$, 
with $t\approx 2.7$\,eV, $a = 1.42$\,\AA, 
$\lambda a \approx 3-4$ \cite{Pereira:2009b,Pellegrino:2010}, and
we disregard bond bending effects, since they are not significant in
this effective description
\cite{Ribeiro:2009}.
Bond deformations are uniform and given to linear order in terms of
the strain tensor
$\ce_{ij}$: $\bm{\bd} = (\bm{1}+\bm{\ce})\cdot\bm{\bd}^0$. 
This approximation was
shown to describe with good accuracy both the threshold deformation
for the Lifshitz and metal-insulator transition at large
deformations \cite{Pereira:2009b,Ni:2009,Choi:2010}, and the behavior
of $t_i(\delta)$ 
when compared to \emph{ab-initio} calculations \cite{Ribeiro:2009}.
Since the hexagonal lattice is elastically isotropic,
$\ce_{ij}$ can be fully parametrized by the
amount of uniaxial strain, $\ce$, its direction $\theta$ with respect
to a zig-zag (\zdir) direction, and Poisson's ratio, $\nu\approx 
0.16$ \cite{Pereira:2009b}. 
Our Cartesian directions are such that a \zdir\ edge coincides with
the $x$ axis (\Fref{fig:Fig-1}). Tension along $\theta=0$ [\zdir] and
$\theta=\pi/2$
[armchair (\adir)] defines representative directions that will be
recalled frequently. We shall disregard the strain-induced
modification of the reciprocal lattice, as it is not relevant for the
optical absorption. Since here we are not interested in large
deformations, we further approximate  
$t_i \equiv t(\delta_i) \approx t - t \lambda (\delta_i - a)$.

The low energy Hamiltonian appropriate for optical processes below
the UV can be derived conventionally \cite{RMP:2009}, by
expanding \eqref{eq:Hamiltonian}
around the \emph{shifted} Dirac points: $\bk =
\bk_D + \bq$ ($q\ll k_D$). For arbitrary $t_i$ the
nonequivalent Dirac points lie at $\bk_{D,\zeta} = \zeta
(A-B,\,A+B)$, with
$ A=\frac{1}{\sqrt{3}a} \cos^{-1}\frac{t_1^2-t_2^2-t_3^2}{2t_2 t_3}, $
$ B=\frac{1}{3a} \cos^{-1}\frac{t_3^2-t_1^2-t_2^2}{2t_1 t_2} $,
and $\zeta=\pm1$ is the valley index, labeling the two nonequivalent
Dirac points. In their vicinity, the electronic
dispersion is given by
\begin{multline}
  E^2_\bq  \simeq
  \frac{9}{4} q_y^2a^2 t_2^2
  +\frac{3\sqrt{3}}{2}q_xq_ya^2 (t_3^2-t_1^2) \\
  +\frac{3}{4}q_x^2a^2 (2t_1^2-t_2^2+2t_3^2)
  + \Ocal  (q^3)
  \label{eq:Eq-general}
  .
\end{multline}
The Fermi surface is consequently an ellipse, with principal axes
that will be rotated with respect to $\Ocal x y$ in the general
situation. \Eqref{eq:Eq-general} can be cast compactly as $E^2_\bq =
\hbar^2v_F^2 \times \bq.\bm{\eta}^2.\bq$, where $\hbar v_F
\equiv 3ta / 2$. Diagonalization of $\bm{\eta}^2$ yields the principal
velocities $v_\pm^2=v_F^2\eta^2_\pm$, with 
\begin{multline}
  \eta^2_\pm =
  \frac{t_1^2+t_2^2+t_3^2}{3t^2} 
  \\ \pm \frac{2}{3t^2}
  \sqrt{t_1^4+t_2^4+t_3^4-t_1^2t_2^2-t_2^2t_3^2-t_1^2t_3^2}
  ,
\end{multline}
and the principal directions:
\begin{equation}
  \tan\varphi_\pm =
  \frac{t_3^2-t_1^2}{\sqrt{3}(\eta_\pm^2t^2-t_2^2)}
  .
\end{equation}
These directions define the slow/longitudinal ($-,l$) and
fast/transverse ($+,t$) directions of
strained graphene, insofar as the direction-dependent Fermi velocity
in the elliptical Fermi surface is smallest along the one, and largest
along the other. 
In the principal coordinate system, the effective Hamiltonian of
valley $\zeta=\pm$ reads
\begin{equation}
  H_{\zeta} \simeq 
    \zeta v_F \tau_1\eta_l p_l + v_F \tau_2\eta_t p_t
  \label{eq:H-eff-strain}
  ,
\end{equation}
where $\tau_{1,2}$ are Pauli matrices acting on the $(A,B)$
sublattice space. 
Coupling to a light wave described by the physical vector potential 
$\bm{A}(t)=\bm{A_0} \exp(-i\omega t)$ is achieved by the minimal
substitution $\bp\to\bp+e\bm{A}$ in \Eqref{eq:H-eff-strain}. The
frequency-dependent conductivity is extracted from the linear response
to $\bm{A}_0$.

For future reference, the anisotropy parameters will be expressed
directly in terms of the longitudinal deformation $\ce$, to first
order. For general tension along $\theta$ with respect to $\Ocal x$ we
have
\begin{subequations}\label{eq:aniso-explicit}
\begin{gather}
  \eta_{t} \simeq 1 + a\lambda\nu\ce,\quad
  \eta_{l} \simeq 1 - a\lambda\ce 
  \label{eq:aniso-explicit-etas}
  ,\\
  \delta \bk_{D,\zeta} \simeq
    \zeta \lambda\ce\frac{1+\nu}{2}(\cos2\theta,-\sin2\theta)
  \label{eq:aniso-explicit-Dirac}
  ,\\
  \tan\varphi_{l} \simeq
    \Bigl[1 - a\lambda\ce\frac{1+\nu}{8} (1+2\cos4\theta)
    \Bigr]\tan\theta
  ,\\
  \tan\varphi_t \simeq
  \Bigl[1 + a\lambda\ce\frac{1+\sigma}{8} (1+2\cos 4\theta)
  \Bigr]\tan\bigl(\theta+\frac{\pi}{2} \bigr)
  .
  \label{eq:aniso-vs-strain}
\end{gather}
\end{subequations}
As intuition dictates, the slow/longitudinal axis is coincident with
the tension axis ($\varphi\approx\theta$). From
\eqref{eq:H-eff-strain} the strain-induced correction to the density
of states (DOS) in the vicinity of the Dirac points follows 
immediately: $\rho(E)\simeq \rho^\text{iso}(E) / (\eta_l \eta_y)$,
where the isotropic DOS reads
$\rho^\text{iso}(E)=2|E|/(\pi\hbar^2v_F^2)$.
Notice that, even though the effective longitudinal (transverse)
velocity increases (decreases), the net effect in the DOS is always a
slope enhancement, because 
$\eta_l \eta_t \approx 1 - a\lambda(1-\nu)\ce <1$. This means that
strain modifies the Fermi energy and/or electron density.

%
%
\section{Conductivity of Strained Graphene}
Kubo's form of the frequency dependent conductivity reads
\begin{equation}
  \sigma_{\alpha\beta}(\omega)
    = \frac{ig_sq^2}{A\omega} \sum_{\bk\lambda\lambda'}
      \frac{
        v^{\bk\lambda\lambda'}_\alpha
        v^{\bk\lambda'\lambda}_\beta
        (n_{\bk\lambda}-n_{\bk\lambda'})
      }{
        \hbar\omega-(\e_{\bk,\lambda'} -\e_{\bk,\lambda})+i0^+
      }
  \label{eq:sigma-Kubo}
  ,
\end{equation}
where $g_s=2$ is the spin degeneracy, $A$ the total area,
$\lambda,\lambda'=\pm$,
and $v^{\bk\lambda\lambda'}_\alpha \equiv
\langle\bm{k},\lambda\vert v_{\alpha}\vert\bm{k},\lambda'\rangle$ are
the matrix elements of the velocity operator
$v_\alpha=i\hbar^{-1}[H,x_\alpha]$ in the momentum eigenbasis. For
the general tight-binding Hamiltonian \eqref{eq:Hamiltonian}, these
matrix elements read explicitly
\begin{multline}
  \langle\bk,\lambda\vert\bm{v}\vert\bk',\lambda'\rangle = 
  -\frac{\lambda}{\hbar}\delta_{\bk,\bk'}\delta_{\lambda\lambda'}
  \sum_\alpha t_\alpha \bd_\alpha \sin\left[\bk.\bn_\alpha
  -\theta_\bk\right]\\
  -\frac{i\lambda}{\hbar}\delta_{\bk,\bk'}\left(1-\delta_{
  \lambda\lambda'}\right)\sum_\alpha t_\alpha\bd_\alpha
  \cos\left[\bk.\bn_{\alpha}-\theta_\bk\right]
  \label{eq:v-matrix}
  ,
\end{multline}
with 
$\theta_{\bk} = \arg \bigl[\sum_i t_i \exp(i\bk.\bn_i)\bigr]$.
Contributions with $\lambda=\lambda'$ are associated with
\emph{intra}-band transitions, and $\lambda\ne\lambda'$ with
\emph{inter}-band, direct transitions. In this form the conductivity
can be directly calculated at the tight-binding level. In order to
proceed fully analytically we work in the Dirac approximation 
\eqref{eq:H-eff-strain}, and consider a clean system at
zero temperature, thus retaining only the inter-band
contribution. In the principal coordinate system defined
by the slow/longitudinal and fast/transverse axes the relevant matrix
elements become:
\begin{subequations}\label{eq:v-matrix-Dirac}
\begin{align}
  |v^{\bk,\lambda,-\lambda}_l|^2 & \approx
    v_F^2\eta_l^2\sin^2\theta_{\bk_D}
  ,\\
  |v^{\bk,\lambda,-\lambda}_t|^2 &\approx
    v_F^2\,\eta_t^2\cos^2\theta_{\bk_D} 
  ,\\
  v^{\bk,\lambda,-\lambda}_l v^{\bk,-\lambda,\lambda}_t &\approx
    \zeta\, v_F^2\tfrac{\eta_l\eta_t}{2} \sin(2\theta_{\bk_D}) 
  .
\end{align}
\end{subequations}
The anisotropy is explicitly encoded both in the parameters
$\eta_{\pm}$, and in the phase $\theta_{\bk_D}$.

From here and \Eqref{eq:sigma-Kubo} it is straightforward to obtain
the frequency dependent
conductivity $\sigma_{\alpha\beta} = \sigma_{\alpha\beta}^\prime + i
\sigma_{\alpha\beta}^{\prime\prime}$. Its real part reads
\begin{equation}
  \sigma_{ll}^\prime(\omega) \simeq  
  \frac{\eta_l}{\eta_t} \times \sigma_0 \times
  \Bigl[
    f\bigl(-\tfrac{\hbar\omega}{2}-\mu\bigr)
    - f\bigl(\tfrac{\hbar\omega}{2}-\mu\bigr)
  \Bigr] 
  \label{eq:Re-sigma}
\end{equation}
for the longitudinal conductivity, and
$\sigma_{tt}^\prime(\omega) = (\eta_t/\eta_l)^2\,\sigma_{ll}
^\prime(\omega)$ for the transverse component.
The isotropic (and universal) value is
$\sigma_0=e^2/(4\hbar)$, and $f(x)$ represents  the Fermi-Dirac
occupation function. The imaginary part, $\sigma''(\omega)$, reads
\begin{equation}
  \sigma_{ll}^{\prime\prime}(\omega) \simeq 
  \frac{\eta_l}{\eta_t} \times \frac{\sigma_0}{\pi}\times
  \log\biggl|\frac{2|\mu|-\omega}{2|\mu|+\omega}\biggr|
  \label{eq:Im-sigma}
\end{equation}
when $T=0$, and $\sigma_{tt}^{\prime\prime}(\omega) =
(\eta_t/\eta_l)^2\,\sigma_{ll}^{\prime\prime}$.
The off-diagonal components $\sigma_{lt}$ are zero,
as one expects from symmetry and the absence of magnetic fields. This
result
shows that $\sigma^\prime_{ii}(\omega)$ is still constant
within the region of validity of the Dirac approximation,
and for $2\mu \lesssim \hbar\omega \ll T$,
being only renormalized by the
anisotropy factors $\eta_{\pm}$. The degree of anisotropy is
controlled by $\eta_l/\eta_t$, which is a sensible result
because
$\sigma_{ii}(\omega)\propto |v^{\bk,\lambda,-\lambda}
_i|^2$, and the ratio reflects the quotient between
the Fermi velocities along the principal strain directions.
From Eqs.~\eqref{eq:Re-sigma} and \eqref{eq:Im-sigma}, and 
recalling the expressions in \Eqref{eq:aniso-explicit-etas}, we
can express the strain induced corrections to the full isotropic
conductivity,
$\sigma(\omega)=\sigma^\prime(\omega)+i\sigma^{\prime\prime}(\omega)$,
in linear order in the deformation as
\begin{equation}
  \sigma_{ll,tt} (\omega)
  \simeq 
  \sigma^\text{iso} (\omega) \times
  \bigl(1 \mp 2 |\delta k_D| a \bigr)
  \label{eq:sigma-strain}
  ,
\end{equation}
where $\sigma^\text{iso} (\omega)$ represents the full
frequency-dependent conductivity in the absence of strain.
For example, tension along \zdir\ ($\theta=0$) yields
a decrease in $\sigma_{xx}$, and an increase of the same magnitude for
 $\sigma_{yy}$. 
Substitution of the material parameters applicable to free-standing
graphene in \eqref{eq:sigma-strain} and
\eqref{eq:aniso-explicit-Dirac} yields an anisotropy factor 
$(\sigma_t - \sigma_l)/\sigma^\text{iso} = 4 |\delta k_D| a \sim
8\ce$. This is sufficiently marked to be detectable by conventional
optical means, like absorption in the visible or IR, which we discuss
below.

\begin{figure*}[tb]
  \centering
  \includegraphics[width=0.9\textwidth]{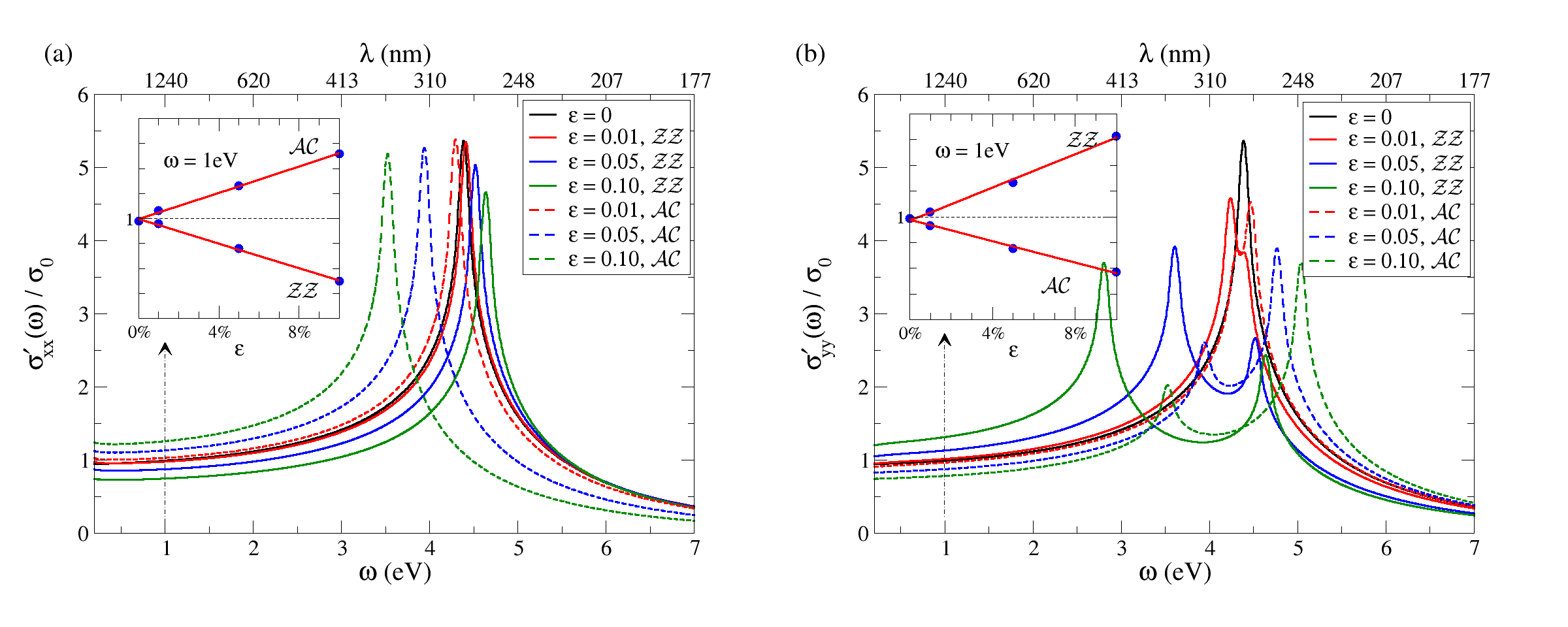}
  \caption{
    Real part of the optical conductivity $\sigma_{xx}$ (a) and 
    $\sigma_{yy}$ (b), calculated
    \emph{ab-initio} for graphene under uniaxial strain along
    \zdir\ ($\theta=0$) and \adir\ ($\theta=\pi/2$).
    The sharp absorption peaks are associated with transitions
    resonant with the van~Hove singularities.
    In panel (a) the peak is resonant with the transition at 
    $M_{1,3}$ in the BZ, whereas in panel (b) the double peak
    structure derives from the splitting between transitions at
    $M_{1,3}$ and $M_2$.
    Both insets show the variation of $\sigma_{ii}^\prime$ with
    strain at $\omega=1\,$eV, as obtained \emph{ab-initio}. The 
    straight lines are linear fits, which yield slopes of    
    $-2.45$ (\zdir) and $2.64$ (\adir) for the inset of panel (a),
    and $3.22$ (\zdir) and $-2.08$ (\adir) for the inset of panel
    (b).
  }
  \label{fig:Fig-2}
\end{figure*}

%
%
\section{Ab-Initio Optical Conductivity}
Given the approximations used, it is legitimate to question the
validity of the analytical corrections written
in \Eqref{eq:sigma-strain}. To that end, we have
extracted the optical conductivity of uniaxially strained graphene
from first-principles as well.
Density Functional Theory (DFT) calculations were performed with the
\emph{ab-initio} spin-density functional code \textsc{aimpro}
\cite{Rayson:2008}. We used the GGA in the scheme
of Perdew, Burke, and Ernzerhof \cite{Perdew:1996}.
Core states were accounted for by dual-space separable
pseudopotentials \cite{Hartwigsen:1998}.
The valence states are expanded over a set of $s$, $p$, and $d$-like
localized, atom-centered Gaussian functions.
The BZ was sampled according to the
scheme proposed by Monkhorst-Pack \cite{Monkhorst:1998}. We used a
convergence-tested grid of $20\times 20\times \times1$ points for
the self-consistent calculations. In equilibrium we obtained the
optimized lattice parameter $a = 1.42$\,\AA. 

Uniaxial strain was applied with relaxation as described earlier
in Ref.~\onlinecite{Ribeiro:2009}. For each strained configuration we 
extracted the dielectric function within the long-wavelength dipole
approximation \cite{Fall:2008}, and from it the optical conductivity.
The results for $\sigma^\prime_{xx}(\omega)$ and
$\sigma^\prime_{yy}(\omega)$ so obtained are shown in
\Fref{fig:Fig-2}, for representative uniaxial strains applied along
\zdir\ ($\theta=0$) and \adir\ ($\theta=\pi/2$).

The main panel of \Fref{fig:Fig-2}(a) shows the renormalization of the
real part of $\sigma_{xx}$, which corresponds to $\sigma_{ll}$ for
\zdir\ and $\sigma_{tt}$ for \adir.
It is clear that for visible and IR frequencies the
conductivity remains roughly constant in $\omega$, but its magnitude
depends on the amount of strain. The strain dependence is shown in
detail in the inset, at $\omega=1\,$eV. 
The perfect linearity in $\ce$ of the \emph{ab-initio} results up to
at least $\ce \approx 10\%$ shows that the analytical result of
\Eqref{eq:sigma-strain} is indeed quite dependable for a wide range of
stretching. We must point out, however, that the linear slopes quoted
in \Fref{fig:Fig-2} cannot immediately be used to extract the
bandstructure parameter $\lambda$ using \Eqref{eq:sigma-strain}.
That is because the \emph{ab-initio} calculation naturally includes
the relaxation and deformation of the lattice, which, as advanced in
the very beginning, we chose to ignore, not to encumber the
discussion by more complicated expressions which do not change the
main results. 

Taking the lattice deformation into account leads to a
correction of the expansion \eqref{eq:Eq-general} of the
dispersion around the Dirac points. This leads to a
renormalization of entries in the matrix $\bm{\eta}^2$. It 
can be shown straightforwardly that such changes amount
to replacing $\lambda a\to\lambda a - 1$ in $\bm{\eta}^2$, and,
consequently, in every ensuing result. In this way, the
\emph{ab-initio} slopes quoted in \Fref{fig:Fig-2}
correspond to     
$\lambda \simeq 2.42\,\text{\AA}^{-1}$ (\zdir) and 
$\lambda \simeq 2.56\,\text{\AA}^{-1}$ (\adir) 
for the inset of panel (a), and 
$\lambda \simeq 2.97\,\text{\AA}^{-1}$ (\zdir) and 
$\lambda \simeq 2.17\,\text{\AA}^{-1}$ (\adir) 
for the inset of panel (b). This rather nicely agrees with the
estimates quoted earlier that put $\lambda a\sim3-4$.

%
%
\section{van~Hove Singularities}
In the UV band, $\sigma(\omega)$ is dominated by direct transitions
between the $M$ points in the BZ, which coincide with VHS in
the electronic dispersion. From the point
of view of \eqref{eq:Hamiltonian}, these transitions occur
at momenta $\bk$ coinciding precisely with
$M_{1,3}=(\pm\pi/\sqrt{3},\,\pi/3)$ and
$M_2=(2\pi/3,\,0)$ (see \Fref{fig:Fig-1} for our convention
regarding the $M$ points). The corresponding resonant frequencies are
therefore given by
$\omega_{Mi} = 2 |E(\bk=M_i)|$, 
and read 
$\omega_{M1} = 2|t_1+t_2-t_3|$, 
$\omega_{M2} = 2|t_1-t_2+t_3|$, and
$\omega_{M3} = 2|-t_1+t_2+t_3|$.
In the most general situation $t_1\ne t_2\ne t_3$ the VHS
split.
Linearly expanding the hoppings in the magnitude of uniaxial strain, 
and defining $\omega_{Mi}\approx2t+ta\lambda\ce\Delta\omega_{Mi}$,
such splitting acquires the explicit form
\begin{subequations}\label{eq:M-splitting}
\begin{align}
  \Delta\omega_{M1,3} & \simeq
  \nu-1+(1+\nu)\bigl(\cos2\theta\pm\sqrt{3}\sin2\theta\bigr)
  ,\\
  \Delta\omega_{M2} & \simeq
  \nu-1-2(1+\nu)\cos2\theta
  .
\end{align}
\end{subequations}
But, as is obvious from the \emph{ab-initio} results
in \Fref{fig:Fig-2}, this splitting is not always visible in the
absorption spectrum. That is because the velocity matrix elements
\eqref{eq:v-matrix} impose a modulation of the strength associated
with these transitions. In order to extract this effect we need to
abandon the Dirac approximation
(\ref{eq:H-eff-strain},\ref{eq:v-matrix-Dirac}), and work with the
full-tight binding bandstructure and matrix elements. In
\eqref{eq:v-matrix} take, for example, the matrix elements at
$\bk=M_2$, which read%
\begin{subequations}\label{eq:v-M2}
\begin{align}
  |v_x^{M_2}|^2 & =
    \frac{3a^2}{4\hbar^2}(t_1-t_3)^2
  ,\\
  |v_y^{M_2}|^2 & =
    \frac{1a^2}{4\hbar^2}(t_1+2t_2+t_3)^2
  ,\\
  v_x^{M_2}v_y^{M_2} & =
    -\frac{\sqrt{3}a^2}{4\hbar^2}(t_1+2t_2+t_3)(t_1-t_3)
  .
\end{align}
\end{subequations}
When expanded in powers of strain, $|v_x^{M_2}|^2$ will be zero to
linear order, irrespective of the strain direction. This suppresses
the $M_2$ singularity in $\sigma'(\omega)$ leaving only the peaks
related to $M_{1,3}$, whose associated matrix elements are
finite. Since the geometry chosen in \Fref{fig:Fig-2} considers
only $\theta=0$ and $\pi/2$, $M_1$ and $M_3$ are
still degenerate, as per \eqref{eq:M-splitting}, and thus only one
peak should survive in $\sigma^\prime_{xx}$, precisely as seen
\emph{ab-initio} in \Fref{fig:Fig-2}(a).
On the other hand, there is no such selection rule arising from the
velocity matrix elements when computing $\sigma^\prime_{yy}$, which
allows the splitting between $M_2$ and $M_{1,3}$ to be observed, as
\Fref{fig:Fig-2}(b) clearly demonstrates. 
Eqs.~\eqref{eq:M-splitting} also explain why in
\Fref{fig:Fig-2}(a) the strain-induced shift 
of the peak is less pronounced for tension along \zdir\
($\theta=0$), than \adir\ ($\theta=\pi/2$): 
$\Delta\omega_{M1}(\zdir) = \nu\ \Delta\omega_{M1}(\adir)$. And
similarly, Eqs.~\eqref{eq:v-M2} account for the different
relative intensity of the peaks associated with $M_2$ and $M_{1,3}$ in
\Fref{fig:Fig-2}(b).

It is important to stress here that these results are not tied
to a particular coordinate system. We can easily express
Eqs.~\eqref{eq:v-M2} in any coordinate system, rotated by $\varphi$
with respect to $\Ocal x y$, and conclude that the longitudinal
matrix element $|v_{x'}^{M_i}|^2$ always vanishes when $\varphi$
coincides with a \zdir\ direction. For example, 
$|v_{x'}^{M_2}|^2$ vanishes for $\varphi=0$,
$|v_{x'}^{M_3}|^2$ vanishes for $\varphi=\pi/3$, and
$|v_{x'}^{M_1}|^2$ vanishes for $\varphi=2\pi/3$. Suppression of a
van~Hove peak in the longitudinal conductivity therefore singles out
one of the \zdir\ directions of the lattice.
This is consistent and explains the numerical calculations of
Ref.~\onlinecite{Pellegrino:2010}, and has an important
consequence: by monitoring the splitting in the absorption peaks
associated with VHS, and by inspecting the selection rule just
described, \emph{one can measure simultaneously: the magnitude of
strain, its direction, and the direction of the underlying lattice
with respect to the laboratory coordinate system}.
This provides an optical means to perform the same measurements
that have been made by exploring the splitting of the $G$ peak in the
Raman spectrum of strained graphene
\cite{Ferrari:2009,Hone:2009}.

\begin{figure}[tb]
  \centering
  \includegraphics[width=0.45\textwidth]{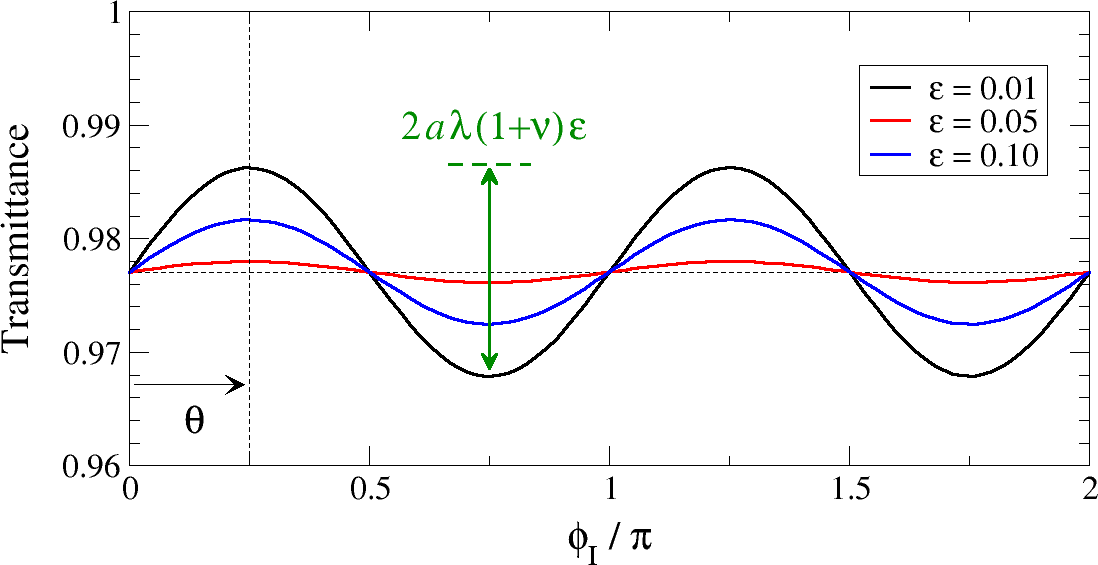}
  \caption{
    Illustration of the behavior of the transmittance as a function
    of incident polarization measured in the laboratory frame.
    According to \eqref{eq:Transmittance}, the phase shift $\theta$
    defines the direction of strain, and the amplitude its magnitude.
  }
  \label{fig:Fig-3}
\end{figure}

%
%
\section{Transparency and Dichroism}
The asymmetry \eqref{eq:sigma-strain} in the conductivity
tensor will result in a certain degree of dichroism, as 
the absorbance of linearly polarized light will depend on the
polarization direction with respect to the slow/fast axes. Treating
graphene as a 2D conducting sheet, and solving the associated Fresnel
equations, we can extract the degree of polarization rotation for
normal incidence on graphene in vacuum. In the visible and IR where
\Eqref{eq:sigma-strain} stands that would be
\begin{equation}
  \frac{\tan\phi_T}{\tan\phi_I} =
    \frac{2+c\mu_0\sigma_{ll}}{2+c\mu_{0}\sigma_{tt}}
    \approx 
    1-4|\delta k_D|a \frac{c\mu_0\sigma_0}{2+c\mu_0\sigma_0}
  \label{eq:Polarization}
  ,
\end{equation}
where $\phi_{T,I}$ are the transmitted and incident
polarizations measured with respect to the slow/longitudinal
axis, and $c\mu_0\sigma_0=\pi\alpha\approx 0.02$. Likewise, the
transmittance for linear polarization becomes
\begin{equation}
  \Tcal \approx 1 - \pi\alpha\bigl[1-2|\delta k_D|a\,\cos2\phi_I\bigr]
  \label{eq:Transmittance}
\end{equation}
An important consequence of this periodic modulation is that it
allows a direct determination of both the strain direction and its
magnitude, as follows. In some laboratory coordinates
\eqref{eq:Transmittance} is transformed by making
$\phi_I\to\phi_I-\varphi_l$. The amplitude of $\Tcal$ as a function of
polarization direction determines the amount of strain, while the
phase shift $\varphi_l$ fixes the direction. This is illustrated in
\Fref{fig:Fig-3}.
The corrections to both polarization and transmittance are weighted
by $\pi\alpha$, and will be necessarily small.
But, one the one hand, the modulation amplitude is roughly 
$\sim 8\ce\pi\alpha$ ($0.16\ce$) and transmittance can be routinely
measured within $0.1\%$ of precision. On the other hand, the
transmittance of multilayer graphene is, to a great extent, cumulative
\cite{Nair:2008,Kuzmenko:2008}. This implies that the same results
apply for multilayer graphene, provided one replaces $\pi\alpha\to
N\pi\alpha$ in
\eqref{eq:Transmittance}, where $N$ accounts for the number of
graphene layers. The effect is thus naturally enhanced in
multilayers, as it is in the vicinity of the VHS (or any resonance,
for that matter).
Another interesting application is that, if strain can be controlled
with precision, an expression like \eqref{eq:Transmittance} allows,
by means of a simple optical experiment, the measurement of the
bandstructure parameter $\lambda = d\log t/dr$, whose knowledge is
crucial for the characterization of all strain-induced effects on the
bandstructure.

%
%
\section{Discussion}
The photoelastic effect in undoped graphene has been quantified, and
shown to possess features that might be appealing in the development
of atomically thin optical elements. One of such characteristics is
the frequency independent response in a very large frequency range,
which remains valid when the system is anisotropically strained. This 
constancy and predictability is a relevant feature for broadband
applications. The degree of anisotropy induced by strain is
determined by how much the Dirac point is displaced from its position
at $K/K'$ in the BZ: $|\delta k_D|$ \eqref{eq:aniso-explicit-Dirac}.
This opens several possibilities, such as: monitoring optical
absorption as a function of strain to characterize the band
parameters of graphene; or monitoring the transmittance as a function
of incident polarization in order to measure the magnitude and
direction of strain in graphene devices. This last example could find
applications in completely passive, transparent, strain sensors.

Even if the magnitude of the photoelastic effect in graphene at
visible or IR frequencies is limited by its small natural absorption
of $\pi\alpha\sim 2\%$, it is nonetheless significant for an
atomically thin membrane. Additional versatility is provided by the
fact that the
effect can be naturally amplified by stacking multilayers, or 
that the optical absorption can be radically affected by 
electronic doping as well, on account of Pauli blocking
\cite{Ando:2002,Basov:2008}.

A determination of the lattice orientation cannot be made
from the optical absorption at low energies, simply due to the
isotropy of the Dirac dispersion. By contrast, the electronic
dispersion at energies close to the VHS fully reflects the symmetry
of the underlying lattice. Therefore, by measuring the optical
response at frequencies resonant with the van~Hove transitions, and
analyzing the strain-induced splitting and selection rules of the
absorption peaks, one can extract the amount of strain, its direction,
and the lattice orientation.

Another consequence of the strain-induced tunability of the VHS is
that it can have a significant import in current efforts to elevate
the Fermi level of graphene up to the VHS
\cite{Barbaros:2010,Kim:2010,Ye:2010}. The ability to achieve this,
and thereby dramatically increase the electronic DOS, is expected to
facilitate many-body instabilities and the establishment of correlated
phases, such as superconductivity, or charge/spin density waves
\cite{Gonzalez:2008,Rotenberg:2010}. Strain engineering of
graphene can, in this respect, work as a facilitator and provide
added tunability. 
For example, from \eqref{eq:M-splitting} it follows that the
energy of the VHS can be reduced by uniaxial strain to values as low
as $E/t \simeq 1-(3+\nu)a\lambda\ce/2$. Using $\lambda a \sim 3$ for
estimate purposes we can write $E/t \simeq 1-5\ce$, so that the
reduction for 10\% strain can be as much as 50\% in the position of
the VHS. On top of this we would need to include excitonic
corrections that are known to renormalize the VHS further down
in energy \cite{Louie:2009}, and which we neglect in our treatment.

Finally, the photoelastic effect is the basis of many optical and
mechano-optical devices and applications at the macro-scale. The
characteristics of graphene in this respect, which we just surfaced,
might provide a valuable route towards the downscaling of those
concepts to the realm of atomically thin optical elements, and their
application at the nanoscale.

%
%
\acknowledgments
AHCN acknowledges the partial support of the U.S. DOE under grant
DE-FG02-08ER46512, and ONR grant MURI N00014-09-1-1063.

%
%
\bibliographystyle{apsrev}
\bibliography{graphene_optical}

\end{document}